\documentclass[aps,showkeys,showpacs,nofootinbib,floatfix]{revtex4}
\usepackage{amssymb}
\usepackage{amsmath}
\usepackage{graphicx}
\usepackage{enumitem}

\def\({\left(}
\def\){\right)}
\def\[{\left[}
\def\]{\right]}

\begin{document}

\title{Tight $H_0$ constraint from galaxy redshfit surveys:
combining baryon acoustic osillation measurements and Alcock-Paczynski test}
\author{Xue Zhang$^{1}$ \footnote{zhangxue@itp.ac.cn},
Qing-Guo Huang$^{1,2,3}$ \footnote{huangqg@itp.ac.cn}
and Xiao-Dong Li$^{4}$
\footnote{corresponding author: lixiaod25@mail.sysu.edu.cn}}
\affiliation{$^1$ CAS Key Laboratory of Theoretical Physics,\\ Institute of Theoretical Physics, \\Chinese Academy of Sciences, Beijing 100190, China\\
$^2$ School of Physical Sciences, \\University of Chinese Academy of Sciences,\\ No. 19A Yuquan Road, Beijing 100049, China\\
$^3$ Synergetic Innovation Center for Quantum Effects and Applications, Hunan Normal University, Changsha 410081, China\\
$^4$ School of Physics and Astronomy, Sun Yat-Sen University, Guangzhou 510297, P. R. China}
\date{\today}

\begin{abstract}
We report a tight Hubble constant constraint $67.78^{+1.21}_{-1.86}$ km s$^{-1}$ Mpc$^{-1}$ (2.26\% precision)
derived from galaxy redshift surveys.
We combine the BAO measurements from 6dFGS, the SDSS DR7 main galaxies, the BOSS DR12 galaxies, and eBOSS DR14 quasars,
and also apply the tomographic Alcock-Paczynski (AP) method to the BOSS DR12 galaxies,
to place constraints on $H_0$ in the spatially flat $\Lambda$CDM framework.
Our result is fully consistent with the CMB constraints from Planck,
but in $2.58\sigma$ tension with local measurements of Riess et al. 2016.
Compared with the BAO alone constraint, the BAO+AP combined result reduces the error bar by 32\%.
This shows the strong power of the tomographic AP method in extracting cosmological information from galaxy redshift surveys.
\end{abstract}

\maketitle


\section{Introduction}

The Hubble constant $H_0$ measures the expansion rate of the present universe.
Its value is closely related to the components of the universe.
Precise determination of $H_0$ has crucial meaning for modern cosmology.

The value of $H_0$ can be directly obtained in local universe measurements,
or indirectly derived via cosmological observations such as the Cosmic Microwave Background (CMB) experiments.
Riess et al. 2016 \cite{Riess:2016jrr} reported a local determination of $H_0=73.24 \pm 1.74$ km s$^{-1}$ Mpc$^{-1}$ (2.4\% precision)
from Cepheids and type Ia supernovae (SNIa)
\footnote{See \cite{Freedman:2010xv,Freedman:2017yms} for recent reviews on astronomical methods of $H_0$ measurements and their significance in cosmology.}.
On the other hand, assuming a flat $\Lambda$CDM model, the nine-year Wilkinson Microwave Anisotropy Probe (WMAP9)
reported a 3\% precision determination of $H_0=70.0 \pm 2.2$ km s$^{-1}$ Mpc$^{-1}$ \cite{Hinshaw:2012aka},
and the Planck survey reported $H_0=67.27 \pm 0.66$ km s$^{-1}$ Mpc$^{-1}$ (1.0\%  precision) \cite{Ade:2015xua}.
There exist a $>3\sigma$ tension between the local measurement and the Planck result.

The tension could be due to unknown systematics in the experiments,
or more interesting, arises because the $\Lambda$CDM model is not good enough
to describe our universe \cite{Wyman:2013lza,Wang:2015wga,Pourtsidou:2016ico,Qing-Guo:2016ykt,DiValentino:2016hlg,Zhao:2017cud,Sola:2017znb,Chen:2017ayg}.
In this situation, having a third method which can determine $H_0$ independently will be very important \cite{Chen:2016uno}.
The galaxy redshift surveys, which map the matter distribution of the low redshift universe
and place independent constraints on the cosmic expansion history,
can provide robust and relatively tight constraints on $H_0$ \cite{Cheng:2014kja}.
Combining the baryon acoustic oscillation (BAO) measurements from current surveys,
Wang et al. 2017 \cite{Wang:2017yfu} reported $H_0=69.13 \pm 2.34$ km s$^{-1}$ (3.38\% precision).

Recently, a novel method was developed to measure cosmic expansion history from the galaxy redshift survey data
via the Alcock-Paczynski (AP) effect \cite{Li:2014ttl,Li:2015jra},
i.e. shape distortion of the large scale structure (LSS) due to incorrectly assumed cosmological parameters \cite{Alcock:1979mp}.
The method focus on the redshift dependence of AP effect to overcome the serious contamination coming from the redshift space distortions (RSD),
and enable extraction of information on relatively small clustering scales.
Li et al. 2016 \cite{Li:2016wbl} (hereafter Li16) applied this method to the
SDSS (Sloan Digital Sky Survey) BOSS (Baryon Oscillation Spectroscopic Survey) DR12 galaxies,
and found that the tomographic AP test leads to tight cosmological constraints competitive with the mainstream cosmological probes
such as SNIa, BAO, and CMB.
Combing the AP method to the other probes improves the constraints on $\Omega_m$ and $w$ by about $35\%$.
So it would be interesting to see whether the Hubble constant constraints can be improved by including the AP method.

In this work we combine the BAO and AP methods to place constraints on $H_0$ from galaxy redshift surveys in the spatially flat $\Lambda$CDM cosmology.
The paper is organized as follows.
In section 2, we will introduce the model, data, and methodology used in this work.
In section 3, we present our main results, and show that
the inclusion of AP method leads to $H_0=67.78^{+1.21}_{-1.86}$ km s$^{-1}$ Mpc$^{-1}$ (2.26\% precision). The conclusion is given in section 4.

\section{Data and methodology}

Throughout our analysis we assume the spatially flat $\Lambda$CDM model to describe our universe.
The angular diameter distance and Hubble parameter take the forms of
\begin{equation}
D_A (z) = \frac{1}{1+z} \int^z_0 \frac{d z'}{H(z')},\ H(z) = H_0 \sqrt{\Omega_r (1+z)^4 + \Omega_m (1+z)^3 + 1-\Omega_r-\Omega_m}.
\end{equation}
The volume-averaged effective distance is
\begin{equation}
D_V (z) \equiv \[ (1+z)^2 D_A^2 (z) \frac{z}{H(z)} \] ^{1/3}.
\end{equation}
Here $\Omega_m$ is the present energy density of matter,
and $\Omega_r = \Omega_{\gamma}(1 + 0.2271 N_\text{eff}) = 4.16 \times 10^{-5}h^{-2}$
is the present energy density of radiation.
We take the photon density $\Omega_{\gamma}h^2 = 2.46 \times 10^{-5}$
and effective number of neutrino species $N_\text{eff} = 3.046$.

\subsection{BAO datasets and methodology}

We use the isotropic BAO measurements from the 6dFGS survey \cite{Beutler:2011hx}, the SDSS main galaxy sample (MGS) \cite{Ross:2014qpa},
the nine anisotropic BAO measurements from BOSS DR12 \cite{Zhao:2016das,Wang:2016wjr},
and isotropic measurement from the eBOSS DR14 \cite{Ata:2017dya}.
Their values and effective redshifts are summarized in Table \ref{tab:BAO}.

For the theoretical value of sound horizon at redshift $z_d$, we use the formula of
\begin{equation}
r_d \equiv r_s (z_d) = \frac{1}{\sqrt{3}} \int^{\frac{1}{1+z_d}}_0 \frac{da}{a^2 H(a) \sqrt{1 + \frac{3 \Omega_b}{4 \Omega_\gamma} a}}.
\end{equation}
where \cite{Eisenstein:1997ik}
\begin{eqnarray}
z_d &=& \frac{1291(\Omega_m h^2)^{0.251}}{1 + 0.659\(\Omega h^2\)^{0.828}} + \[1 + b_1 (\Omega_b h^2)^{b_2}\],\\
b_1 &=& 0.313(\Omega_m h^2)^{-0.419}\[1 + 0.607(\Omega_m h^2)^{0.674}\],\ \
b_2 = 0.238(\Omega_m h^2)^{0.223}.
\end{eqnarray}
We adopt the Planck result $\Omega_b h^2 = 0.02225$  \cite{Ade:2015xua}.

\begin{table}
\centering
\begin{tabular}{c c c c c}
  \hline
  \hline
  $z_\text{eff}$ & Variable & Value (Mpc) & Experiment & Reference \\
  \hline
  0.106 & $D_V$ & $456 \pm 27$ & 6dFGS & \cite{Beutler:2011hx} \\
  \hline
  0.15 & $D_V$ & $(664 \pm 25)(r_d/r_{d,\text{fid}})$ & SDSS DR7 MGS & \cite{Ross:2014qpa} \\
  \hline
  0.31 & $D_A$ & $ (6.29 \pm 0.14 ) r_d$ \\
  0.36 & $D_A$ & $ (7.09 \pm 0.16 ) r_d$ \\
  0.40 & $D_A$ & $ (7.70 \pm 0.16 ) r_d$ \\
  0.44 & $D_A$ & $ (8.20 \pm 0.13 ) r_d$ \\
  0.48 & $D_A$ & $ (8.64 \pm 0.11 ) r_d$ & BOSS DR12 (9zbin) & \cite{Zhao:2016das,Wang:2016wjr} \\
  0.52 & $D_A$ & $ (8.90 \pm 0.12 ) r_d$ \\
  0.56 & $D_A$ & $ (9.16 \pm 0.14 ) r_d$ \\
  0.59 & $D_A$ & $ (9.45 \pm 0.17 ) r_d$ \\
  0.64 & $D_A$ & $ (9.62 \pm 0.22 ) r_d$ \\
  \hline
  1.52 & $D_V$ & $(3843 \pm 147)(r_d/r_{d,\text{fid}})$ & eBOSS DR14 & \cite{Ata:2017dya}  \\
  \hline
  \hline
\end{tabular}
\caption{BAO distance measurements used in this work.}
\label{tab:BAO}
\end{table}

%

\subsection{AP datasets and methodology}

Refs. \cite{Li:2014ttl,Li:2015jra} proposed to use the redshift dependence of the Alcock-Paczynski (AP) effect as a cosmological probe.
The apparent anisotropy in the observed galaxy sample arise from two main sources,
the RSD effect due to the galaxy peculiar velocities \cite{FOG,Kaiser,Kwan2012,Zhang2013,Zheng2013,Li:2016bis},
and the geometric distortion when incorrect cosmological models are assumed for
transforming redshift to comoving distance, known as the AP effect.
They found that anisotropies produced by RSD are, although large,
maintaining a nearly uniform magnitude over a large range of redshift,
while the degree of anisotropies from the AP effect significantly varies with redshift.
So they focus on the redshift dependence of the anisotropic clustering of galaxies,
which is less affected by RSD contamination, but still sensitive to cosmological parameters.

Li16 applies the methodology to the BOSS DR12 galaxies.
The 361\,759 LOWZ galaxies at $0.15<z <0.43$ and
771\,567 CMASS galaxies at $0.43< z < 0.693$ are split into six redshift bins (three bins in LOWZ and three in CMASS);
at each bin, the integrated 2-point correlation function (2PCF) is computed
\begin{equation}
 \xi_{\Delta s} (\mu) \equiv \int_{s_{\rm min}}^{s_{\rm max}} \xi (s,\mu)\ ds.
\end{equation}
Here $\xi$ is measured as a function of $s$ and $\mu$, where $s$ is the distance between the galaxy pair,
and $\mu=\cos(\theta)$ with $\theta$ being the angle between the line joining the pair of galaxies and
the line of sight (LOS) direction to the target galaxy.

The range of integration, chosen as $s_{\rm min}=6$ $h^{-1}$Mpc and $s_{\rm max}=40$ $h^{-1}$Mpc,
controls the clustering scale investigated by the analysis.
By reducing the RSD effect via focusing on the redshift dependence of anisotropy,
the complicated modeling of RSD and non-linear clustering is avoided.
As a result, it becomes possible to use the galaxy clustering down to 6 $h^{-1}$Mpc.
This is a major advance in deriving cosmological constraints from small clustering scales,
where there are a lot of independent structures, and the amount of information is enormous.

To mitigate this systematic uncertainty from galaxy bias and clustering strength,
the analysis only relies on the shape of $\xi_{\Delta s}(\mu)$,
\begin{equation}\label{eq:norm}
 \hat\xi_{\Delta s}(\mu) \equiv \frac{\xi_{\Delta s}(\mu)}{\int_{0}^{\mu_{\rm max}}\xi_{\Delta s}(\mu)\ d\mu}.
\end{equation}
A cut $\mu<\mu_{\rm max}$ is imposed to reduce the fiber collision and FOG effects which are strong toward the LOS ($\mu\rightarrow1$) direction.

The redshift evolution of anisotropy between galaxies in the $i$th and $j$th redshift bins is quantified by
\begin{equation} \label{eq:deltahatxi}
\delta \hat\xi_{\Delta s}(z_i,z_j,\mu)\ \equiv\ \hat\xi_{\Delta s}(z_i,\mu) - \hat\xi_{\Delta s}(z_j,\mu).
\end{equation}
The systematics of $\delta \hat\xi_{\Delta s}$ (hereafter $\delta\hat\xi_{\Delta s, \rm sys}$) mainly comes from the redshift evolution of RSD effect.
Refs. \cite{Li:2014ttl,Li:2015jra} found the RSD effect creates small redshift dependence in $\hat \xi_{\Delta s}(\mu)$,
mainly due to the structure growth and the selection effect
(different galaxy bias at different redshifts).
In the analysis, we use the mock galaxy sample from the Horizon Run 4 \cite{Kim:2015yma} N-body simulations to estimate this systematics.
The galaxy assignment scheme applied to HR4 is very successful in modeling the
large scale Kaiser effect and the small scale FOG effect in nonlinear regions \cite{Hong:2016hsd}.

Li16 chose the {\it first redshift bin} as the reference and compare the measurements in higher redshift
bins with that in the first.
Then the redshift evolution of anisotropy in the whole sample is quantified by
\begin{equation}\label{eq:chisq1}
\chi_{\rm AP}^2\equiv \sum_{i=2}^{6} \sum_{j_1=1}^{n_{\mu}} \sum_{j_2=1}^{n_{\mu}} {\bf p}(z_i,\mu_{j_1}) ({\bf Cov}_{i}^{-1})_{j_1,j_2}  {\bf p}(z_i,\mu_{j_2}),
\end{equation}
where $n_{\mu}$ denotes the binning number of $\hat\xi_{\Delta s}$, and ${\bf p}(z_i,\mu_{j})$ is defined as
\begin{eqnarray}\label{eq:bfp}
 {\bf p}(z_i,\mu_{j}) \equiv&\ \delta \hat\xi_{\Delta s}(z_i,z_1,\mu_j) - \delta \hat\xi_{\Delta s, \rm sys}(z_i,z_1,\mu_j).
\end{eqnarray}
The covariance matrix ${\bf Cov}_i$ is the covariance matrix of ${\bf p}$ estimated from the 2,000
MultiDark-Patchy mock surveys of BOSS DR12 \cite{Reid:2015gra}.
In wrong cosmologies, the AP effect produces large evolution of clustering anisotropy,
thus would be disfavored according to Eq. \eqref{eq:chisq1}.

Li16 tested the robustness of the method
and found the derived cosmological constraints insensitive to the adopted options
within the range of $s_{\rm min}=2-8\ h^{-1}$ Mpc,  $s_{\rm max}=30-50\ h^{-1}$ Mpc, $\mu_{\rm max}=0.85-0.99$,
and number of binning  $n_\mu=6-40$.
The results are also robust against choices of mocks adopted for systematic correction and covariance estimation.
In this work we will not re-conduct the above tests, which are not the main topic of this paper.
We take the options of $s_{\rm min}=6$  $h^{-1}$ Mpc,
$s_{\rm max}=40\ h^{-1}$ Mpc, $\mu_{\rm max}=0.97$, and $n_\mu=20-25$
\footnote{Our $n_\mu$ is smaller than the default choice of Li16 (where they more ambitiously chose $n_\mu=20-35$);
this weakens the constraints by a little bit, but increases the robustness of the results,
and also reduces the noise in the likelihood.}.

\subsection{Combining AP and BAO}

The BAO method uses the BAO feature in the clustering of galaxies on scales of 100-150 $h^{-1}$Mpc,
created by the oscillation of the baryon-photon plasma in the early Universe.
Measuring this scale in 1D or 2D then yields measurements of $D_V$ or $D_A$ and $H$ at some representative redshifts.
As a comparison, the tomographic AP method mainly utilizes the redshift dependence of $D_A(z) H(z)$.
Its clustering scale $6-40\ h^{-1}$Mpc is much smaller than BAO scale,
and the correlation between galaxy clustering in the scales probed by the two methods should be rather independent.
In Appendix A, we conduce a simple check and found that the information explored by the two methods is fairly independent.
So in the cosmological analysis we can easily combine them without worrying about their correlation.

We constrain $\Omega_m$ and $H_0$ through Bayesian analysis \cite{Christensen:2001gj},
which derives the probability distribution function (PDF) of some parameters $\bf \theta$ (=($\Omega_m,H_0$) in this paper)
given observational data $\bf D$,
according to Bayes' theorem.
Assuming flat priors for $\Omega_m$ and $H_0$,
and approximate the PDF by a likelihood function $\mathcal{L}$ satisfying  $-2 \ln \mathcal{L}=\chi^2$, we have
\begin{equation}
 P({\bf \theta}|{\bf D}) \propto \mathcal{L} \propto \exp\left[-\frac{\chi^2}{2}\right].
\end{equation}
We use the {\texttt {COSMOMC}} software \cite{Lewis:2002ah}
to obtain the Markov Chain Monte Carlo (MCMC) samples of $\theta$ following the PDF of $P({\bf \theta}|{\bf D})$
and derive constraints on $\Omega_m$ and $H_0$.

\section{Results}

\begin{figure}[!htbp]
\centering
\includegraphics[width=.49\textwidth, height=.5\textwidth]{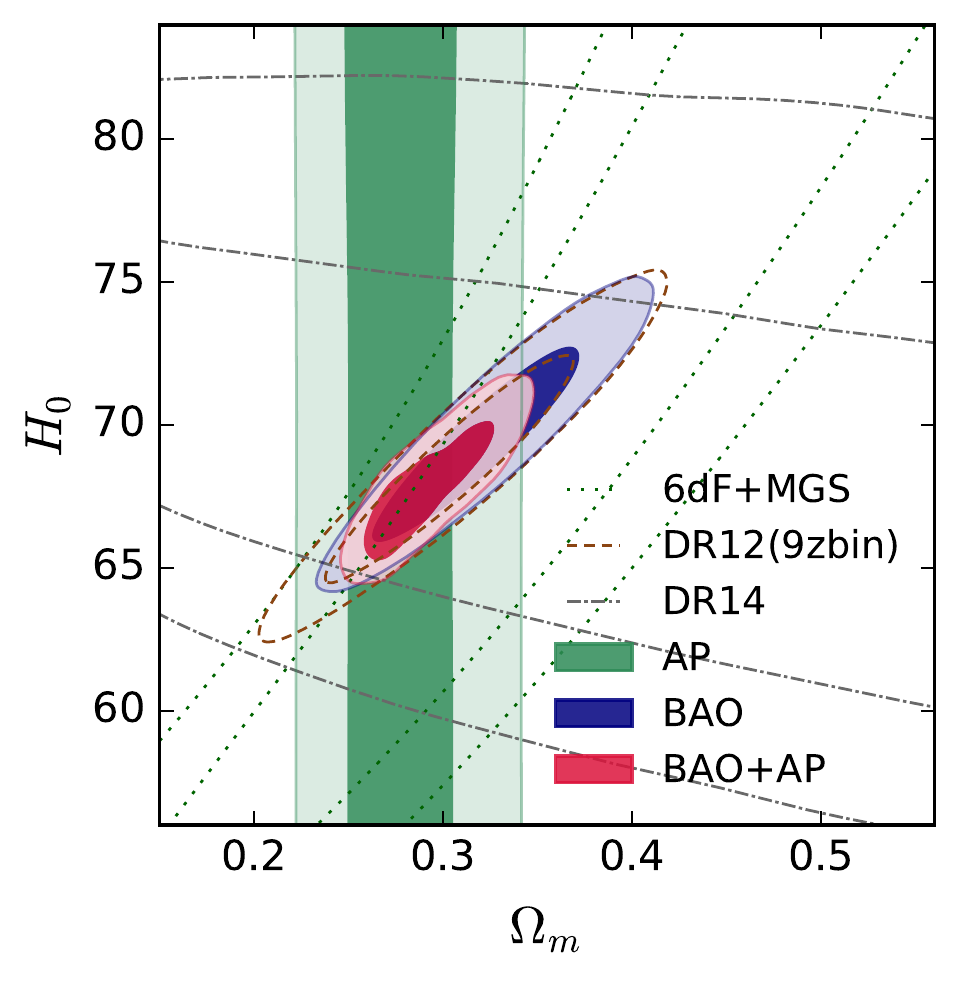}
\hfill
\includegraphics[width=.49\textwidth, height=.5\textwidth]{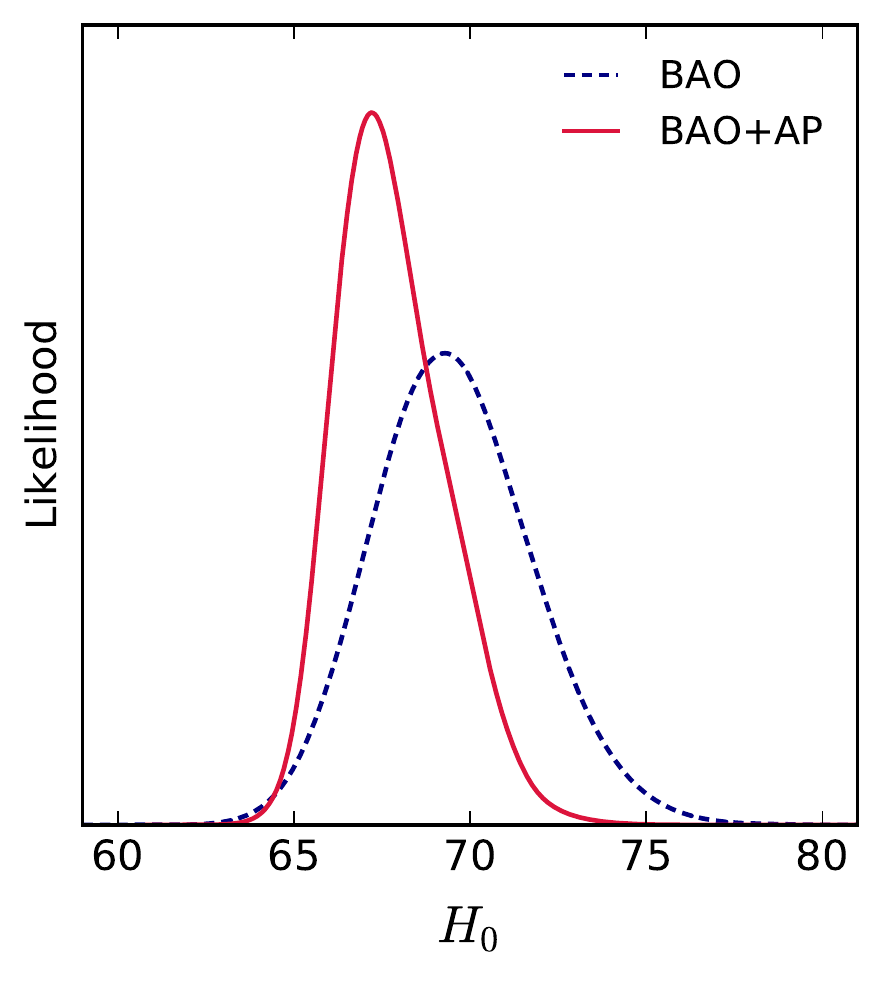}
\caption{Left panel:
Likelihood contours (68.3\% and 95.4\%) of $H_0$ derived from various data sets,
where BAO represents a combination of 6dF+MGS+DR12 (9zbin)+DR14.
Adding AP tightens the constraint on $\Omega_m$ and reduces the contour area by 50\%.
Right panel: Likelihoods of $H_0$ from BAO (red solid) and BAO+AP (blue dotted).
Adding AP reduces the uncertainly by 32\%.}
\label{fig:combine}
\end{figure}

We present the likelihood contours of $H_0$ and $\Omega_m$ in the left panel of FIG. \ref{fig:combine}.
Using the BAO datasets we can simultaneously constrain these two parameters.
Here we plot the results using datasets of 6dF+MGS, DR12, DR14, and all combined (labeled by BAO).
The shapes and positions of contours are well consistent with Wang et al. 2017 \cite{Wang:2017yfu}.
We find the all combined BAO result leads to
\begin{equation}
\Omega_m = 0.32^{+0.03}_{-0.04},\ H_0=69.51^{+2.08}_{-2.44}\ \rm km s^{-1} Mpc^{-1}\ (BAO\ alone).
\end{equation}
The two parameters are strongly degenerated with each other.


The AP method is sensitive to the shape of clustering and does not care about their absolute size.
It can not directly constrain $H_0$, which controls the absolute scale of the structures.
Incorrect value of $\Omega_m$ would change the cosmic expansion history non-uniformly and introduce an AP effect,
so it can be effectively constrained by our method,
and we get
\begin{equation}
\Omega_m = 0.278 \pm 0.022\ (\rm AP\ alone).
\end{equation}
This constraint is tighter than the BAO constraint.

The BAO+AP combined result is plotted in the red region of FIG. \ref{fig:combine}.
Although the AP method alone can not place constraints on $H_0$,
it assists the BAO method by tightens the constraint on $\Omega_m$
and breaks the degeneracy between the two parameters.
We get
\begin{equation}
\Omega_m = 0.290\pm 0.023,\ H_0=67.78^{+1.21}_{-1.86}\rm \ km\ s^{-1}\ Mpc^{-1}\ (BAO+AP).
\end{equation}
After adding the AP method, the contour area is reduced by 50\%,
and the uncertainty of $H_0$ is reduced by 32\%.


The right panel of FIG. \ref{fig:combine}
shows the marginalized likelihood distribution of $H_0$.
Clearly, adding AP significantly reduces the statistical uncertainty.
The central value is shifted to smaller values
and becomes closer to the CMB results.



\begin{figure}[!htbp]
\centering
\includegraphics[width=.55\textwidth]{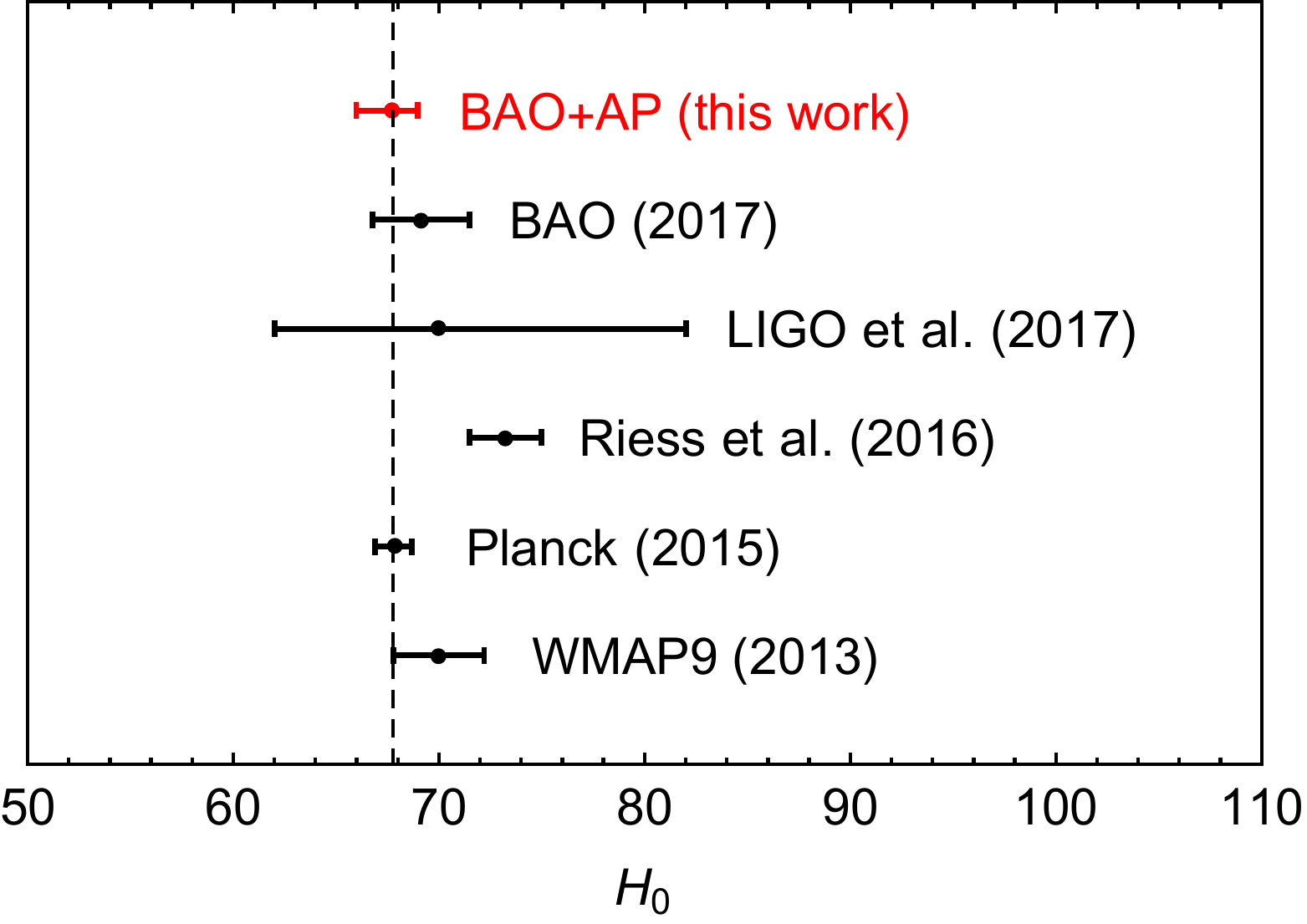}
\caption{Comparison of current $H_0$ measurements.
In this work we get $H_0=67.78^{+1.21}_{-1.86}$ km s$^{-1}$ Mpc$^{-1}$,
which is fully consistent with CMB results but in $2.58$ tension with the local measurement results of Riess et al. 2016.}
\label{fig:errorbar}
\end{figure}

The $H_0$ measurements from different methods are summarized in FIG. \ref{fig:errorbar}.
The BAO+AP result is fully consistent with the Planck CMB result,
while in $2.58 \sigma$ tension with the local result of Riess et al. 2016.
But it is still not large enough to conclude
a detection of discrepancy between the two sets of results.

On 17 August 2017, the Advanced LIGO and Virgo detectors \cite{Abbott:2017xzu}
observed the strong signal of GW170817 from the merger of a binary neutron-star system,
and determine the Hubble constant to be about $70.0^{+12.0}_{-8.0}$ km s$^{-1}$ Mpc$^{-1}$.
This result is completely independent of CMB and local measurements.
Although the current error bar is too large,
we expect much better constraints from future gravitational wave experiments. 

\section{Summary and discussion}
\label{sd}

In this paper, we determine the Hubble constant using galaxy redshift survey datasets within the flat $\Lambda$CDM cosmology.
Combining the BAO and AP methods we found that the Hubble constant can be precisely determined as
$67.78^{+1.21}_{-1.86}$ km $\rm s^{-1} \rm Mpc^{-1}$ (2.26 \% precision).
The result is consistent with the Planck result $67.27 \pm 0.66$ km s$^{-1}$ Mpc$^{-1}$,
and in $2.58 \sigma$ tension with the  Riess et al. 2016 result $73.24 \pm 1.74$ km s$^{-1}$ Mpc$^{-1}$.

Throughout the analysis, we assume a flat $\Lambda$CDM cosmology model.
The derived results heavily relies on this assumption.
The results would change or be weakened if one considers alternatives or extensions of $\Lambda$CDM model.


Compared with the BAO alone case, the constraints on $H_0$ is improved by 32\% after adding AP into the analysis.
This shows the strong power of the new AP method in extracting information from galaxy clusterings.
It would be worthy investigating the constraints on other parameters using the tomographic AP method,
and we will work on this issue in future investigations.

The major caveat of the AP analysis is that the RSD effect is estimated from the HR4 mock survey samples
created in a particular cosmology of $\Omega_m=0.26$ $\Lambda$CDM model. 
There would be a systematic bias in the estimation if this adopted cosmology is different from the truth. 
We believe that this is not a serious problem in our analysis due to the following reasons.
First of all, the simulation cosmology is within $1.5\sigma$ of $\Omega_m$ constraints, so the analysis is self-consistent;
secondly, Li et al. 2014 \cite{Li:2014ttl} shows that the redshift dependence of RSD is not sensitive to cosmological parameters;
finally, Li et al. 2017 \cite{Li:2017nzs} found that discarding the systematics correction changes the derived constraints by only $<0.5\sigma$,
implying that the systematics of this method is not a serious problem for the precision of current redshift surveys.
But it is necessary to develop a model-independent method of systematic correction
to ensure that the tomographic AP method can be safely applied to future galaxy surveys.


We find the current $H_0$ constraint derived from the galaxy redshift surveys is already tighter than the local measurement from Cepheids and SNIa.
With the progressive development of LSS experiments and statistical methods,
we expect much more precise determination of $H_0$ in the near future.

\vspace{5mm}
\noindent {\bf Acknowledgments}

We acknowledge the use of HPC Cluster of ITP-CAS.
Xiao-Dong thanks Zhaofeng Kang, Jian Li, Yuting Wang, Yi Wang, Xin Zhang and Gong-bo Zhao for kind helps.
This work is supported by grants from NSFC (grant NO. 11335012, 11575271, 11690021, 11647601), Top-Notch Young Talents Program of China, and partly supported by Key Research Program of Frontier Sciences, CAS.

Funding for SDSS-III has been provided by the Alfred P. Sloan Foundation, the Participating Institutions, the National Science Foundation, and the U.S. Department of Energy Office of Science. The SDSS-III web site is http://www.sdss3.org/.

SDSS-III is managed by the Astrophysical Research Consortium for the Participating Institutions of the SDSS-III Collaboration including the University of Arizona, the Brazilian Participation Group, Brookhaven National Laboratory, Carnegie Mellon University, University of Florida, the French Participation Group, the German Participation Group, Harvard University, the Instituto de Astrofisica de Canarias, the Michigan State/Notre Dame/JINA Participation Group, Johns Hopkins University, Lawrence Berkeley National Laboratory, Max Planck Institute for Astrophysics, Max Planck Institute for Extraterrestrial Physics, New Mexico State University, New York University, Ohio State University, Pennsylvania State University, University of Portsmouth, Princeton University, the Spanish Participation Group, University of Tokyo, University of Utah, Vanderbilt University, University of Virginia, University of Washington, and Yale University.

We acknowledge the use of Horizon Run simulations.

\section{Appendix A. Checking the correlation between the tomographic AP and BAO methods}

As a quick check of their independency,
we tests the correlation between the anisotropic clustering on small and large scales.
We define
\begin{equation}\label{eq:y}
 y_{\rm AP} = \frac{\int_{0}^{0.5} d\mu\ \int_{s_1=6 h^{-1}{\rm Mpc}}^{s_2=40  h^{-1}{\rm Mpc}}\ ds \ \ \xi (s,\mu)}
 {\int_{0.5}^{\mu_{\rm max}} d\mu\ \int_{s_1=6  h^{-1}{\rm Mpc}}^{s_2=40  h^{-1}{\rm Mpc}}\ ds\ \  \xi (s,\mu)},\ \
 y_{\rm BAO} = \frac{\int_{0}^{0.5} d\mu\ \int_{s_1=90  h^{-1}{\rm Mpc}}^{s_2=130  h^{-1}{\rm Mpc}}\ ds\ \  \xi (s,\mu)}
 {\int_{0.5}^{\mu_{\rm max}} d\mu\ \int_{s_1=90  h^{-1}{\rm Mpc}}^{s_2=130  h^{-1}{\rm Mpc}}\ ds\ \  \xi (s,\mu)},
\end{equation}
where $y_{\rm AP}$, $y_{\rm BAO}$ quantify the anisotropy of galaxy clustering on scales of AP and BAO method, respectively.

\begin{figure}[!htbp]
\centering
\includegraphics[width=12cm]{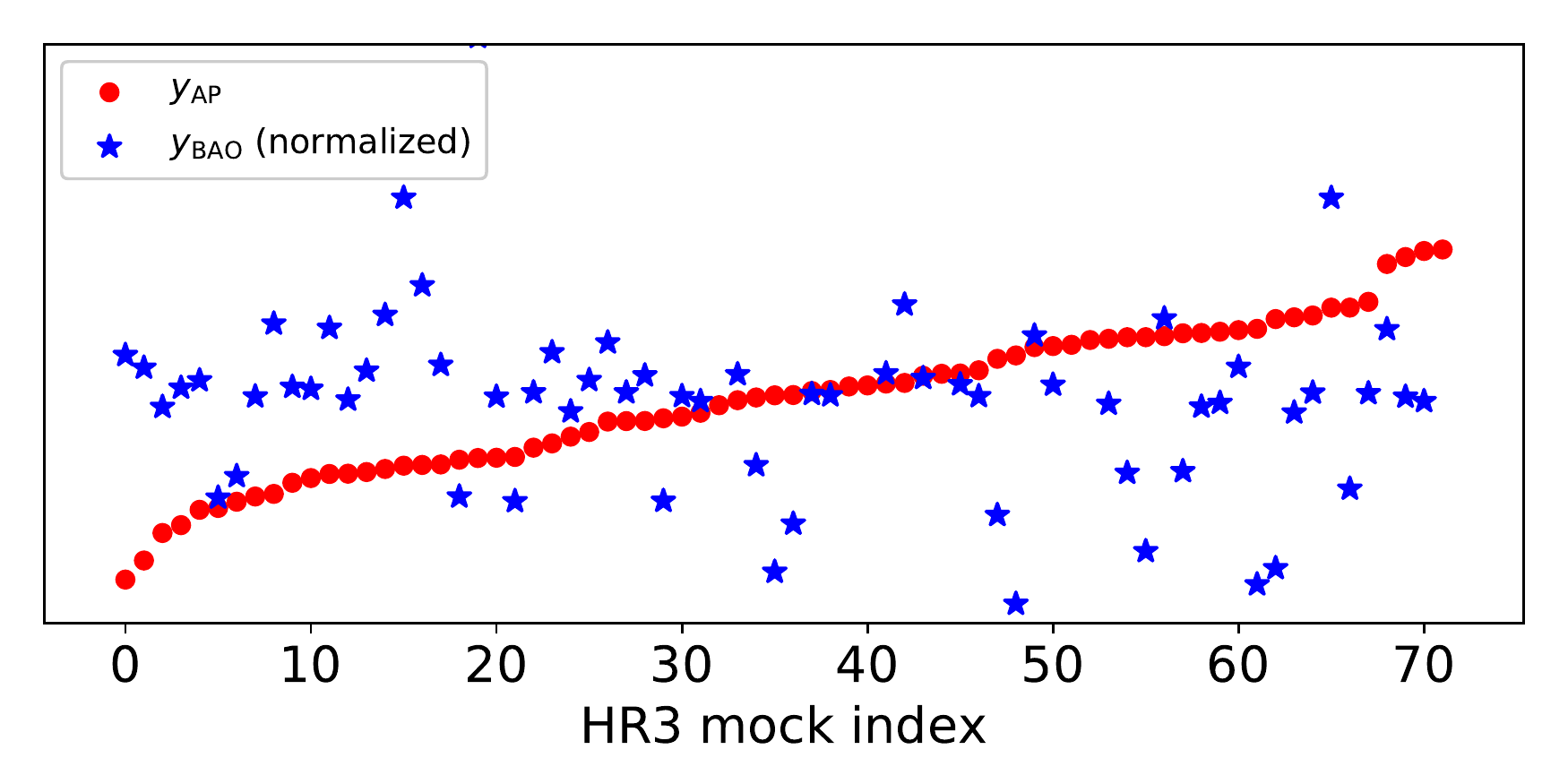}
\caption{Checking the correlation between the AP and BAO methods.
$y_{\rm AP}$ and $y_{\rm BAO}$, as defined in Eq. \ref{eq:y}, characterize the anisotropy
of clustering on the scales probed by the AP and BAO methods.
We measure their values in 72 sets of HR3 mocks and find they are basically uncorrelated.
}
\label{fig:correlate}
\end{figure}

We compute their values in 72 sets of Horizon Run 3 (HR3) CMASS mocks \cite{Kim:2011ab,Li:2016wbl},
and plot the results in Figure \ref{fig:correlate}
\footnote{Here the value of $y_{\rm BAO}$ is normalized to having the similar mean and standard deviation with $y_{\rm AP}$.}.
In case we sort the mocks in order of increasing $y_{\rm AP}$, we see the values of $y_{\rm BAO}$ randomly scattered.
This suggests that they are statistically uncorrelated.
We compute the correlation coefficient of $y_{\rm AP}$ and $y_{\rm BAO}$ and find
\begin{equation}
 r = -0.054 \pm 0.034.
\end{equation}
This result is statistically consistent with no correlation.
So it should be OK to ignore the correlation between AP and BAO in the analysis.

\end{document}